\documentclass[aps,preprint,showpacs]{revtex4}
\usepackage{psfig}
\begin{document}        %
\draft

\title{Comments on "Nonlinear Elastic Response in Solid Helium: Critical Velocity or Strain?"}
\author{Zotin K.-H. Chu} 
\affiliation{Physics Department, Centre for Distribution, Road Xihong, Wulumuqi 830000, China} 
\begin{abstract}
We make comments on  Day {\it et al.}'s [{\it Phys. Rev. Lett.}
104, 075302 (2010)] paper. Our focus is upon the hysteresis loop
in Figure  3 of this paper which  was not closed.

%
\end{abstract}

\maketitle
\bibliographystyle{plain}
Day {\it et al.} recently measured the shear modulus over a wide
frequency range and they then found out, in contrast with the
torsional oscillator behavior, the elastic shear modulus depends
on the magnitude of stress, but not velocity [1]. In [1] they
showed the  hysteresis in the shear modulus (cf. Fig. 3 [1]). In
fact as reported in [1] hysteresis appears when the sample is
cooled below 60 mK and is nearly temperature independent below 45
mK. Day  {\it et al.}  interpreted their (hysteresis) results in
terms of the motion of dislocations which are weakly pinned by
$^3$He impurities but which break away when large stresses are
applied. \newline The present author has other different
explanation than above or [1] and relevant remarks especially
about the nearly closing of the hysteresis loop in Fig. 3 [1].
Firstly, after rearranging the data in [1] ($\sigma=\mu\epsilon$,
cf. Fig. 4 for 2000Hz curve [1]), we can understand there are
plastic flows [2-3] during the experiments. This can be evidenced
in Fig. 1 as the shear stress-strain curve is not linear around
$\epsilon \ge 0.3 \times 10^{-6}$ (yielding occurs). Meanwhile
although the (shear) stress-strain data is isothermal but as the
hysteresis loop in Fig. 3 of [1] is not closed thus there are
adiabatic (shear) stress-strain regime in Fig. 3 of [1]. The
latter can be rearranged and illustrated in Fig. 2. The crucial
note is the adiabatic increase of temperature triggers the thermal
softening phenomenon and reduces the rate of strain hardening (the
shear modulus or (flow) stress can be reduced $17\%$ as shown in
Fig. 1 of [1]). To be concise when the outer layer of solid $^4$He
is subjected to plastic distortion most of the work done reappears
in the form of heat, but a certain proportion remains latent and
is associated with the changes to which plastic distortion give
rise in the physical properties of the solid $^4$He. When the
solid $^4$He is heated all the latent heat must be released before
the melting point is reached, and when it is dissolved the latent
heat must appear as a heat of solution. It seems to the present
author that the identifying of the adiabatic regime is essential
to the possible observation of superfluidity of solid $^4$He in
[1].

\psfig{file=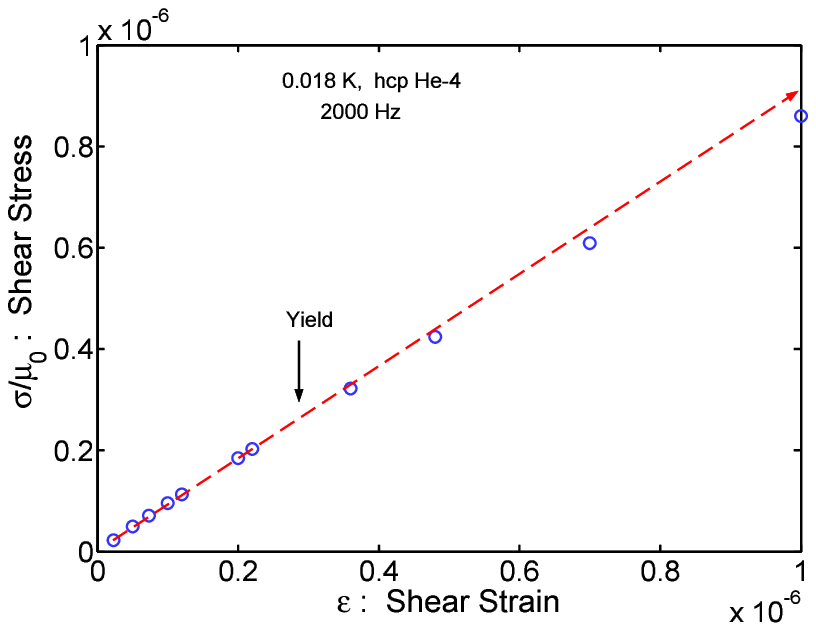,bbllx=0.4cm,bblly=19.5cm,bburx=9cm,bbury=26.8cm,rheight=7cm,rwidth=8.8cm,clip=}
\begin{figure}[h]
FIG. 1. 
Possible plastic flow in Day {\it et al.}'s
measurements (cf. Fig. 4 in [1]). The onset of yielding relates to
the departure from the linear relation (shear stress vs. strain).
\end{figure}
\psfig{file=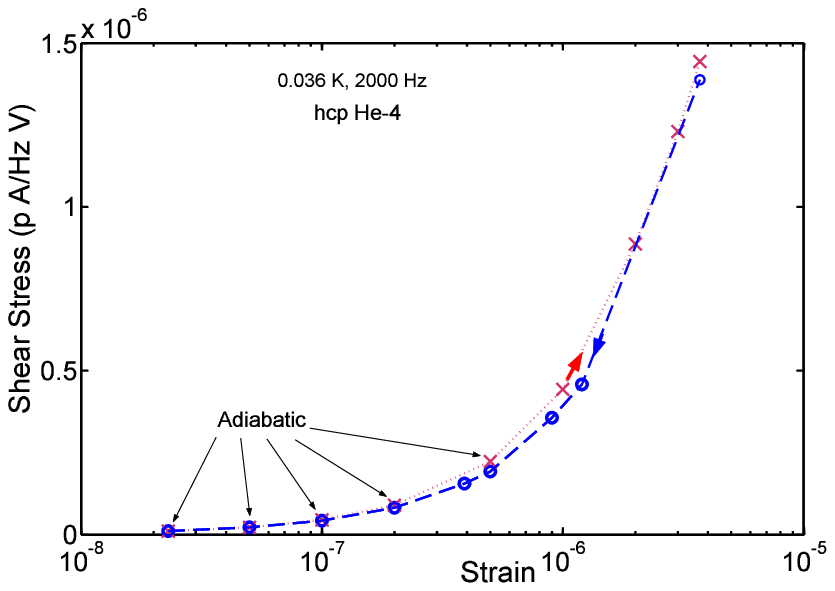,bbllx=-0cm,bblly=19.5cm,bburx=12cm,bbury=26.8cm,rheight=7cm,rwidth=10cm,clip=}

\begin{figure}[h]
FIG. 2. 
Possible adiabatic (shear) stress-strain regime in Day {\it et
al.}'s measurements (cf. Fig. 3 in [1]). In fact the hysteresis
loop is not closed in the isothermal stress-strain data (cf. Fig.
3 in [1]). Arrows follow those in Fig. 3 of [1].
\end{figure}
\end{document}